\documentclass[preprint,12pt]{aastex}

\shorttitle{Intensity Distributions}

\begin{document}

\title{Gamma-Ray Burst Intensity Distributions}
\author{David L. Band\altaffilmark{1}, Jay P. Norris and Jerry T. Bonnell}
\affil{GLAST SSC, Code 661, NASA/Goddard Space Flight
Center, Greenbelt, MD  20771} \altaffiltext{1}{Joint Center
for Astrophysics, Physics Department, University of
Maryland Baltimore County, 1000 Hilltop Circle, Baltimore,
MD 21250} \email{dband@lheapop.gsfc.nasa.gov,
jnorris@lheapop.gsfc.nasa.gov,
jerry@milkyway.gsfc.nasa.gov}

\begin{abstract}
We use the lag-luminosity relation to calculate
self-consistently the redshifts, apparent peak bolometric
luminosities $L_B$, and isotropic energies $E_{\rm iso}$
for a large sample of BATSE gamma-ray bursts.  We consider
two different forms of the lag-luminosity relation; for
both forms the median redshift for our burst database is
1.6. We model the resulting $E_{\rm iso}$ sample with power
law and Gaussian probability distributions without redshift
evolution, both of which are reasonable models.  The power
law model has an index of $\alpha_E=1.76\pm0.05$ (95\%
confidence), where $p(E_{\rm iso}) \propto E_{\rm
iso}^{-\alpha_E}$.  The simple universal jet profile model
suggested but did not require $\alpha_E=2$, and subsequent
physically reasonable refinements to this model permit
greater diversity in $\alpha_E$, as well as deviations from
a power law; therefore our observed $E_{\rm iso}$
probability distribution does not disprove the universal
jet model.
\end{abstract}


\section{Introduction}

The major breakthroughs in the study of gamma-ray bursts of
the past six years---most if not all bursts are
cosmological, bursts do {\it not} have constant peak
luminosity, the fireballs are beamed, many bursts are
associated with supernovae---resulted from the intensive
study of a relatively small number of bursts without regard
for whether these bursts formed a well defined statistical
sample.  However, realizing that for any one burst we are
only sampling an anisotropic radiation pattern from one
direction, we now have to study distributions of burst
properties to reconstruct the bursts' appearance from all
directions. Well-defined burst samples are required to
derive these distributions.

Unfortunately, we do not yet have a large sample of bursts
with spectroscopic redshifts, which in most cases are
required for calculating the intrinsic burst properties.
The two dozen or so bursts with redshifts were detected by
various detectors with different detection thresholds and
energy sensitivities, and the follow-ups that determined
the redshifts depended on the vagaries of weather and
telescope availability. BATSE provided a large burst sample
with well-understood thresholds, but without direct
redshift determinations for most of these bursts. However,
redshifts can be determined indirectly from the
lag-luminosity (Norris, Marani \& Bonnell 2000) or
variability-luminosity (Fenimore \& Ramirez-Ruiz 2000;
Reichart et al. 2001) relations. Here we calculate
self-consistently the redshifts for a large subset of the
BATSE bursts using the lag-luminosity relation.  To
accommodate the problematic burst GRB980425, which is
significantly underluminous if the lag-luminosity relation
is a single power law, Salmonson (2001) and Norris (2002)
proposed that the relation should be a broken power law.
This reduces the luminosity of many long-lag bursts, moving
this population closer to the observer.

We use the resulting redshifts to calculate both the peak
bolometric luminosity $L_B$ and the isotropic energy
$E_{\rm iso}$.  Bursts are thought to radiate
anisotropically, but we sample their radiation field in
only one direction.  Therefore $L_B$ and $E_{\rm iso}$ are
calculated from the observed flux as if the emissions are
isotropic; corrections for the anisotropy are based on
models of the relativistic jets that emit the observed
gamma rays (see Bloom, Frail \& Kulkarni 2003).  $L_B$ and
$E_{\rm iso}$ are both bolometric quantities in the burst
frame.  $L_B$ is the maximum value of the luminosity while
$E_{\rm iso}$ is the total energy emitted over the duration
of the burst (without correcting for the anisotropic
emission).  We consider the isotropic energy $E_{\rm iso}$
to be more fundamental, and therefore model its probability
distribution.

Two models have been proposed for the structure of the
jets. The uniform jet model (Frail et al. 2001; Bloom et
al. 2003) assumes that all jets have a constant surface
energy density $\epsilon$ (energy per solid angle across
the jet) but differing opening angles $\theta_0$ (the angle
between the jet axis and the edge of the jet). The model
makes no predictions for the energy probability
distribution.  On the other hand, in the universal jet
profile model (Rossi, Lazzati, \& Rees 2002; Zhang \&
Meszaros 2002) all jets have the same surface energy
density $\epsilon$ as a function of off-axis angle $\theta$
(the angle from the jet axis), and thus the observed
differences in $L_B$ or $E_{\rm iso}$ result from the angle
$\theta_v$ between the jet axis and the line-of-sight. This
model predicts that the energy probability distribution is
a power law with index $\alpha_E$, where $p(E_{\rm iso})
\propto E_{\rm iso}^{-\alpha_E}$. If $\epsilon \propto
\theta^k$, then $\alpha_E=1-2/k$. Rossi et al. (2002) and
Zhang \& Meszaros (2002) suggested $k=-2$, resulting in
$\alpha_E=2$, to reproduce the distributions observed by
Frail et al. (2001), although this was not a firm
prediction. A Gaussian surface energy density profile
results in $\alpha_E=1$ (Lloyd-Ronning, Dai \& Zhang 2004).
Thus $\alpha_E$ is of order 1--2. Lloyd-Ronning et al.
(2004) showed that allowing the parameters of the jet
profile to vary somewhat from burst to burst---the jet
profile is now only quasi-universal---results both in
variations in the value of $\alpha_E$ and deviations from a
pure power law. While numerical modelling of hypernovae
(Zhang, Woosley, \& MacFadyen 2003) shows that the surface
energy density of the outflows should vary with off-axis
angle, it does not predict the profile; thus the profiles
are fitted empirically to the data.  Consequently, the
observed $E_{\rm iso}$ probability distribution is not a
powerful discriminant between the two jet structure models.

This paper has two goals.  First, we calculate the redshift
of a large sample of bursts self-consistently from the
lag-luminosity relation.  With the redshift and the
observed fluxes and fluences we calculate $L_B$ and $E_{\rm
iso}$. Second, we model the $E_{\rm iso}$ distribution with
two functional forms:  power law and lognormal.  We find
the best fit values for the parameters of each functional
form, and evaluate how well each functional form describes
the data. We also compare the fit of the power law
functional form to the predictions of the universal jet
profile model.

In these calculations we use a cosmology with $H_0 =
70$~km~s$^{-1}$~Mpc$^{-1}$, $\Omega_m = 0.3$ and
$\Omega_\Lambda = 0.7$.  The notation $p(a\,|\,b)$ means
the probability of $a$ given $b$.  Lower case $p$ denotes a
probability density while upper case $P$ (without a
subscript) represents a cumulative probability.  Note that
$a$ and $b$ are propositions that are true or false. A
proposition can be a statement such as ``the energy
distribution can be described with a lognormal functional
form'' or a parameter value (equivalent to the statement
``the parameter value is 5'').

In \S 2 we present the methodology used in this study:
calculating the burst redshifts (\S 2.1); fitting the
energy probability distribution (\S 2.2); and using the
cumulative probability to test the quality of the fit (\S
2.3).  The implementation of this methodology is discussed
in \S 3 and the results are provided by \S 4. Finally, our
conclusions are in \S 5.

\section{Methodology}

\subsection{Calculating Redshifts}

In the absence of a large number of measured spectroscopic
redshifts, we use the lag-luminosity relation (Norris et
al. 2000) to calculate peak luminosities $L_B$, and
subsequently redshifts, for a large burst sample. Assume
that for a set of bursts we have measured the lag $\tau_0$
(s) between the light curves in two energy bands, the peak
photon flux $P_{\rm ph}$ (ph~s$^{-1}$~cm$^{-2}$) in the
photon energy band $E_L$--$E_U$, and the spectral fit for
the time period over which $P_{\rm ph}$ is measured. The
spectral fit ($N_{\rm
ph}(E)$---ph~cm$^{-2}$~s$^{-1}$~keV$^{-1}$) uses the
``Band'' functional form (Band et al. 1993) which is
characterized by an asymptotic low energy power law with
index $\alpha$ ($N_{\rm ph}(E)\propto E^\alpha$), a high
energy power law with index $\beta$ ($N_{\rm ph}(E)\propto
E^\beta$), and a characteristic energy $E_p$ between the
two power laws; for $\beta<-2$ $E_p$ is energy at the peak
of the $E^2 N_{\rm ph}$ (energy per logarithmic frequency
interval) light curve.  This functional form is fitted to
the observed count spectrum, which covers a limited energy
range, but is extrapolated to high and low energy, as
needed. The instrument team is assumed to have converted
the observed peak count rate into the peak photon flux
using the fitted spectral model and the response function.

The lag $\tau_B$ in the burst's frame and the observed lag
$\tau_0$, both between the same energy bands in their
respective frames, are related by time dilation (which
increases the lag) and spectral redshifting (which shifts
the smaller lag at high energy into the observed band). The
first effect is simple and universal---a factor of
$(1+z)^{-1}$. The second depends on the burst's spectral
evolution, and may vary from burst to burst. Thus
\begin{equation}
   \tau_B = \tau_0 g(z) \quad .
\end{equation}
We assume that $g(z)$ is universal, and as a working
assumption we use $g(z)=(1+z)^{c_\tau}$.  Time dilation
contributes $-1$ to $c_\tau$, while the redshifting of
temporal structure with a smaller lag from higher energy
contributes a positive constant (e.g., $\sim 1/3$; Fenimore
et al. 1995 found $\sim 0.4$).

Empirically the lag has been related to the apparent
bolometric peak luminosity $L_B$ (erg~s$^{-1}$; Norris et
al. 2000)
\begin{equation}
L_B = Q(\tau_B) \quad.
\end{equation}

The observed peak bolometric energy flux $F_B$
(erg~cm$^{-2}$~s$^{-1}$) is related to the bolometric peak
luminosity $L_B$ through the redshift, and an assumed
cosmology:
\begin{equation}
F_B = L_B / [4\pi D_L^2] \quad ,
\end{equation}
where $D_L$ is the luminosity distance.  Note that $L_B$ is
the ``isotropic'' peak bolometric luminosity, the peak
luminosity if the observed flux were beamed in all
directions.  If the flux is actually beamed into a solid
angle $\Delta \Omega$, then the actual peak luminosity is
only $\Delta\Omega/4\pi$ of $L_B$. We define
\begin{eqnarray}
\langle E \rangle &=& F_B/P_{\rm ph} \\ \hbox{where} \quad
   P_{\rm ph} &=& \int_{E_L}^{E_U} N_{\rm ph}(E) dE \\
   \hbox{and} \quad F_B &=& \int_0^\infty E N_{\rm ph}(E) dE
\quad ;
\end{eqnarray}
$N_{\rm ph}(E)$ is the photon spectrum.  The photon
spectrum energy flux $F_B$ results from a fit to the
observed spectrum which is available over a limited energy
band from a time bin that includes the peak of the light
curve. Thus the peak bolometric energy flux $F_B$ is
calculated by extrapolating $N_{\rm ph}(E)$ in eq.~6 to
high and low energies.

The resulting implicit equation for redshift,
\begin{equation}
P_{\rm ph} = {{Q\left( \tau_0 g(z) \right)}\over{4 \pi
   D_L^2 \langle E \rangle}} \quad ,
\end{equation}
must be solved for each burst.  The inputs are $\tau_0$,
$P_{\rm ph}$ and $\langle E \rangle$ (which is calculated
from the spectral fit). The functional form of the
lag-luminosity relation $Q$ is calibrated from the small
set of bursts for which $z$ is known. We have an assumed
functional form for $g(z)$ with one unknown constant
$c_\tau$. This constant can be calculated using the
dependence of $\tau$ on energy bands (Fenimore et al.
1995).

These equations can be evaluated for the sensitivity of
calculated quantities such as $z$ and $L_B$ on observables
such as $P_{\rm ph}$ and $\langle E \rangle$ (which, in
turn, depends on the spectral parameters).  Eq.~7 shows
that $z=\phi(P_{\rm ph}\langle E \rangle,\tau_0)$ while
eqs.~3--4 can be combined to give $L_B = P_{\rm ph}\langle
E \rangle \psi(z)$, where $\phi$ and $\psi$ are functions
defined only for this sensitivity analysis and need not be
derived explicitly.  The quantities $P_{\rm ph}$ and
$\langle E \rangle$ appear in these expressions only in the
product $P_{\rm ph}\langle E \rangle$, and consequently
errors in $\langle E \rangle$, resulting from uncertainties
in the spectral parameters, are equivalent to errors in
$P_{\rm ph}$.  As shown by figure~1, these equations give
curves in $z$--$L_B$ space parameterized by $\tau_0$;
larger values of the product $P_{\rm ph}\langle E \rangle$
lie at larger values of $z$ and $L_B$ on the appropriate
curve.

If the spectrum is fairly flat, e.g., $\alpha=-1$ and
$\beta=-2$, then $\langle E \rangle$ does not vary by a
large factor as $E_p$ varies.  The segments plotted on
figure~1 are for $E_p$ ranging between 5~keV and 2~MeV for
$\alpha=-1$ and $\beta=-2$ while holding $P_{\rm ph}$
fixed. However, if the spectrum is more peaked, i.e., is
harder at low energies (larger $\alpha$) or softer at high
energy (smaller $\beta$), then the $\langle E \rangle$
range is larger. This is shown on figure~1 by the diamonds,
triangles and asterisks that mark the range that results as
$E_p$ varies for different sets of $\alpha$ and $\beta$. As
can be seen, cavalier treatment of the spectrum of the peak
flux can result in large errors in $z$ and $L_B$.

Solving eq. 7 provides $z$ for each burst (note that $L_B$
does not appear explicitly in eq. 7). With $P_{\rm ph}$ and
the energy fluence (erg~cm$^{-2}$) we then calculate $L_B$
and $E_{\rm iso}$.  The result is a large database of
observables (lags, $P_{\rm ph}$, energy fluences, spectral
parameters) and derived quantities ($z$, $L_B$, and $E_{\rm
iso}$) with which we can address different questions about
the burst population.  Here we study the distribution of
intensity measures such as $L_B$ and $E_{\rm iso}$.

\subsection{The Intensity Probability Distribution}

We use a Bayesian analysis to investigate intensity
probability distributions, although we argue that a
frequentist analysis will give the same result; Band (2001)
provides a more complete exposition of the methodology.  In
this study we consider the isotropic energy $E_{\rm iso}$
as the relevant intensity measure; the same formulae apply
substituting $L_B$ for $E_{\rm iso}$.

Our burst database provides the isotropic energy $E_{\rm
iso}$, the peak photon flux $P_{\rm ph}$ and the redshift
$z$ for each burst.  The threshold peak flux $P_{\rm
ph,min}$ is known, from which the threshold isotropic
energy $E_{\rm iso,min}$ for each burst is calculated:
$E_{\rm iso,min} = E_{\rm iso} \,P_{\rm ph,min}/P_{\rm
ph}$. The energy probability distribution function is the
normalized probability distribution $p(E_{\rm iso} \,|\,
\vec{a_j}, M_j, I)$ where $\vec{a_j}$ is the set of
parameters that characterize the $j$th model distribution
function represented by $M_j$, and $I$ specifies general
assumptions about our calculation. Thus $M_j$ states that
we are using the $j$th functional form, which has
parameters $\vec{a_j}$.  There are always additional
assumptions upon which the calculation rests, such as the
underlying cosmology or the detector calibration.  The
validity of our analysis depends on these assumptions,
which we represent with the proposition $I$. If the
distribution evolves with redshift, then $z$ should be
included in the list of `givens' in the probability
distribution; here we do not include a redshift dependence.
Thus we do not consider redshift evolution in our analysis.
Because we work with probability distributions that are
normalized at each redshift, our results remain valid for
evolution in the burst rate per comoving volume.

In our calculations we use power law and lognormal energy
probability distributions.  In the power law case the
proposition $M_j=M_{\rm pl}$ is the statement that the
functional form is a power law, and the parameters
$\vec{a_j}$ are the power law index $\alpha_E$ and the low
energy cutoff $E_2$:
\begin{equation}
p(E_{\rm iso}\,|\, \alpha_E, E_2, M_{\rm pl}, I )\propto
E_{\rm iso}^{-\alpha_E} \qquad \hbox{for} \qquad E_{\rm
iso}\ge E_2 \quad .
\end{equation}
For a finite number of bursts the distribution must have a
low energy cutoff $E_2$ if $\alpha_E>1$.  As we will see,
the fit to the power law model is relatively insensitive to
the low energy cutoff $E_2$. Therefore our results are
valid for redshift evolution in the energy scale (i.e., in
$E_2$), but not in the power law index.

In the lognormal case $M_j=M_{\ln}$ states that the
functional form is a lognormal, and the parameters are the
energy centroid $E_{\rm iso,cen}$ and the logarithmic width
$\sigma_E$:
\begin{equation}
p(E_{\rm iso}\,|\, E_{\rm iso,cen}, \sigma_E, M_{\rm ln},
I) = {1\over{\sqrt{2\pi}\sigma_E}} \exp\left[ {{\left(
   \ln \left(E_{\rm iso}\right)-\ln \left(E_{\rm iso,cen}\right)\right)^2}
   \over {2\sigma_E^2}}\right] \quad .
\end{equation}

For the $i$th burst we have $E_{\rm iso,i}$ with threshold
$E_{\rm iso,min,i}$, the set of which constitutes the data
$D$. The observed $E_{\rm iso,i}$ are not drawn from
$p(E_{\rm iso} \,|\, \vec{a_j}, M_j, I)$ but from
\begin{equation}
p(E_{\rm iso} \,|\, E_{\rm iso,min}, \vec{a_j}, M_j, I)
   = {{p(E_{\rm iso} \,|\, \vec{a_j}, M_j, I)
   H(E_{\rm iso} - E_{\rm iso,min})}
   \over {\int_{E_{\rm iso,min}}^\infty dE_{\rm iso} \,
   p(E_{\rm iso} \,|\, \vec{a_j}, M_j, I)}}
\end{equation}
where $H(x)$ is the Heaviside function (1 for positive $x$
and 0 for negative $x$); $E_{\rm iso}$ is drawn from the
observable part of the energy distribution function.

The probability of obtaining the data $D$ given the model
is the `likelihood'
\begin{equation}
   \Lambda_j= p(D\,|\, \vec{a_j}, M_j, I) =
   \prod_{i=1}^{N_B} p(E_{\rm iso,i} \,|\, E_{\rm iso,min,i},
   \vec{a_j}, M_j, I)
\end{equation}
where the product is over the $N_B$ bursts in our database.
In the ``frequentist'' framework best-fit parameters are
typically found by maximizing $\Lambda_j$ through varying
the parameters. Plotting $\Lambda_j$ as a function of
$\vec{a_j}$ around the maximum reveals the range of
acceptable $\vec{a_j}$ values.

However the Bayesian analysis is based on $p(\vec{a_j}
\,|\, D,M_j,I)$, the posterior probability for the
parameters, that is, the probability that the particular
set of $\vec{a_j}$ values is correct given the data. Bayes
theorem gives
\begin{equation}
p(\vec{a_j} \,|\, D,M_j,I) = {{p(D\,|\, \vec{a_j},M_j,I)
   p(\vec{a_j}\,|\, M_j,I)} \over
   {\int d\vec{a_j} \, p(D\,|\, \vec{a_j},M_j,I)
   p(\vec{a_j}\,|\, M_j,I)}} \quad .
\end{equation}
The factor $p(\vec{a_j}\,|\, M_j,I)$ is the prior for
$\vec{a_j}$, constraints on the parameters based on
information available before the new data were acquired.
The factor in the denominator is a normalizing constant.
Thus for parameter determination the Bayesian approach
explicitly factors in additional information and
constraints (e.g., energies must be non-negative and
finite). The expectation value of the parameters is
\begin{equation}
\langle \vec{a_j} \rangle = \int d\vec{a_j} \,\vec{a_j}\,
   p(\vec{a_j}\,|\, D,M_j,I) \quad .
\end{equation}
If $\Xi_j = p(D\,|\, \vec{a_j},M_j,I) p(\vec{a_j}\,|\,
M_j,I)$, the numerator in eq.~12, is sharply peaked, then
the expectation value of the parameters occurs at the peak
of $\Xi_j$. However the posterior probability $p(\vec{a_j}
\,|\, D,M_j,I)$ (or $\Xi_j$) is required to determine the
acceptable parameter range.

As was discussed in Band (2001), reasonable priors for the
parameters of both the power law and lognormal
distributions are constants (when the logarithm of an
energy is treated as the parameter rather than the energy),
and therefore the frequentist likelihood $\Lambda_j$ and
the Bayesian posterior $p(\vec{a_j} \,|\, D,M_j,I)$ are
proportional to each other.  Consequently, while we favor
the Bayesian approach, here the frequentist and Bayesian
analyses are the same. We calculate $\Lambda_j$ as a
function of the parameters $\vec{a_j}$. The maximum of
$\Lambda_j$ gives the best fit parameter values while the
width of $\Lambda_j$ shows the range of acceptable values,
and correlations between the parameter values.

Note that the methodology we use here only provides the
normalized probability distribution, not the normalization
of the distribution (i.e., the burst rate per volume).  The
normalization requires the history of $E_{\rm iso,min}$
during the mission (i.e., not only for the detected
bursts).

\subsection{The Cumulative Probability}

The Bayesian approach does not provide a goodness-of-fit
statistic.  However a frequentist statistic can be derived.
For each burst the cumulative probability is
\begin{equation}
P(>E_{\rm iso,i} \,|\, E_{\rm iso,min,i},\vec{a_j},M_j,I)
   = \int_{E_{\rm iso,i}}^\infty p(E_{\rm iso} \,|\,
   E_{\rm iso,min,i},\vec{a_j},M_j,I)\, dE_{\rm iso}
   \quad .
\end{equation}
If the assumed energy distribution function is an
acceptable characterization of the observations (which
would be the case if the model $M_j$ is correct) and all
the assumptions (included in the proposition $I$) are valid
(e.g., the cosmological model is correct), then the
cumulative probabilities $P(>E_{\rm iso,i})$ for each burst
should be uniformly distributed between 0 and 1, and have
an average value of $\langle P(>E_{\rm iso,i}) \rangle=
1/2\pm (12N_B)^{-1/2}$ for the $N_B$ bursts in the sample.
Note that $(12N_B)^{-1/2}$ is the expected statistical
variance resulting from the size of the sample, and does
not take into account systematic errors.

\section{Implementation}
\subsection{Datasets}

We start with a database for 1438 BATSE bursts that
includes the lags and their uncertainties, the peak flux
$P_{\rm ph}$ over the 50--300~keV band on the 256~ms
timescale, the burst duration $T_{90}$, and hardness ratios
among the 4 BATSE energy channels (30--50, 50--100,
100--300, and 300--2000~keV). Of these 1438 bursts, 1218
have positive lags.

To calculate the average energy $\langle E\rangle$ (see
eq.~4) we use the parameters of the ``Band'' spectrum fits
by Mallozzi et al. (1998) to the 16 channel BATSE ``CONT''
count spectra accumulated over the peak flux time interval
(usually 2.048~s) for the 580 bursts in our database that
are also in the Mallozzi et al. database.  For the 858
bursts that are not in the Mallozzi et al. database we
assume the average low and high energy spectral indices of
$\alpha=-0.8$ and $\beta=-2.3$ found by Preece et al.
(2000).  Because $E_p$, the energy of the peak of the $E^2
N_{\rm ph} (E)\quad(\propto \nu f_\nu)$ curve, and
HR$_{32}$, the 100--300~keV to 50--100~keV hardness ratio,
are strongly correlated, we use the empirical relation
$E_p$ = 240 HR$_{32}^2$~keV (for HR$_{32}\le 2.25$) for the
bursts without spectral fits.  In \S 2.1 we discussed the
sensitivity of $z$ to errors in the spectrum; the gaps in
the available spectral fits introduces a systematic
uncertainty into our analysis.  Figure~2 shows the
resulting scatter plot for $E_p$ vs. HR$_{32}$.

\subsection{The Lag-Luminosity Relation}

Based on less than a dozen bursts, Norris et al. (2000)
found that $L_B\propto \tau_B^{-1.15}$.  However, this
single component relationship predicts a much greater $L_B$
than observed for GRB980425, the burst that appears to
coincide with the supernova SN1998bw. By introducing a
break in the lag-luminosity relation Salmonson (2001) and
Norris (2002) were able to include GRB980425. The resulting
lag-luminosity relation is:
\begin{equation}
L_{51} = 2.18 (\tau_B/0.35\hbox{ s})^{c_L} , \quad c_L=
    -1.15 \hbox{ for $\tau_B \le 0.35$ s, }
    -4.7 \hbox{ for $\tau_B > 0.35$ s.}
\end{equation}
where $L_{51}=L_B/10^{51}$ erg~s$^{-1}$.  In our
calculations we use both the single (i.e., $c_L = -1.15$
for all $\tau_B$) and two component lag-luminosity
relations.

\section{Results}

\subsection{Lag-Luminosity Relation and Redshift}

We first investigate the lag-luminosity relation and then
use it to construct a database of burst redshifts, $L_B$
and $E_{\rm iso}$. Salmonson (2001) and Norris (2002)
introduced a break in the simple power law relation to
include the long lag, very low luminosity GRB980425.  As a
result of this break other long lag bursts in the BATSE
database are assigned low luminosities, and consequently
small distances; Norris (2002) suggested that there may be
a population of nearby low luminosity bursts. Figure~3
demonstrates this:  the solid line is the cumulative
distribution of burst redshifts assuming a single power law
lag-luminosity relation, while the dashed line shows the
distribution for the broken power law relation. Introducing
a break in the power law shifts the low redshift bursts
closer (i.e., to lower redshifts).  Note this break was
introduced only to include GRB980425 in the lag-luminosity
relation; if GRB980425 was not associated with SN1998bw or
if GRB980425 was anomalous, then this break is unnecessary
and unsupported by other data.

Figure~3 also shows that there are few bursts with $z>10$.
These bursts may indeed be at high redshift; finding such
high redshift bursts is one of the goals of the Swift
mission.  However, on theoretical grounds bursts with
$z>17$ are not expected.  Given the dispersion in the
lag-luminosity relation and the errors in determining
$\tau_0$, it is not surprising that some bursts are
assigned redshifts as high as $z=65$; large redshifts
result from high luminosities for bursts with very small
lags, which are particularly difficult to measure
accurately.  Thus the lag-luminosity relation does not give
unphysical results at the high luminosity end. Norris
(2002) placed an upper limit on the luminosity when he
found a large number of very high redshift bursts.

We use the simple single power law lag-luminosity relation
without any cutoffs or limits in the following analysis.

For our database the median redshift is $z_m = 1.58$. As
can be seen from Figure~3, this redshift is greater than
the redshift range where the form of the lag-luminosity
relation makes a difference.

Figure~4 shows the distribution of $L_B$ we calculate (the
scatter plot) along with the threshold of $L_B$ at
different redshifts (the lines). In calculating the
threshold values of $L_B$ we assume a constant $P_{\rm
ph,min} = 0.3$~ph~cm$^{-2}$~s$^{-1}$ and vary $\langle E
\rangle$ (see eq.~ 4). Factor of $\sim 3$ differences in
$\langle E \rangle$ result in significant over- or
under-predictions of the threshold.

Thus we have accomplished the first goal of this paper.
The resulting database is provided online at
http://cossc.gsfc.nasa.gov/analysis/lags/ .

\subsection{The $E_{\rm iso}$ Probability Distribution}

We focus on the probability distribution functions of the
bolometric energy.  To perform quantitative estimates of
distributions as a function of redshift (e.g., of the burst
rate) the threshold for the database must be known, and
therefore studies have often focused on the peak luminosity
$L_B$; $L_B$ is more closely related to the peak flux
$P_{\rm ph}$, which has a relatively sharp instrumental
threshold $P_{\rm ph,min}$. However, to study probability
distributions of intrinsic burst quantities the threshold
for that quantity for each observed burst will suffice.
Thus our database is sufficient to study both the isotropic
energy $E_{\rm iso}$ and the peak luminosity $L_B$
probability distributions.

Figure 5 shows the scatter plot of the isotropic energy
$E_{\rm iso}$.  The region just above the threshold is
underpopulated, suggesting that the thresholds are
underestimated for this database. Therefore, we raise the
threshold to $P_{\rm ph,min}=0.5$~ph~cm$^{-2}$~s$^{-1}$,
raising the median redshift from 1.58 to 1.62, and
decreasing the number of bursts to 1054.
\subsubsection{Power Law $E_{\rm iso}$ Distribution}
This functional form is relevant because it is predicted by
the universal jet profile model. Figure~6 shows the
resulting likelihood surface, which peaks at
$\alpha_E=1.76$ and $\log(E_2)=50.2$.  This value of $E_2$
is the smallest value of $E_{\rm iso}$ in our sample; burst
samples that extend to fainter bursts will probably have
smaller observed values of $E_{\rm iso}$, and consequently
$E_2$ is most likely smaller.  The contour plot indicates
that the determination of $\alpha_E$ is independent of the
value of $E_2$.  Since the likelihood peaks at the same
$\alpha_E$ at any given $E_2$, our result holds even if the
energy scale evolves with redshift.  The 95\% confidence
region centered on $\alpha_E=1.76$ has a half width of
0.05.

As discussed in \S 1, the universal jet profile model,
especially with physically reasonable refinements, does not
predict definitively a value of $\alpha_E$ that can then be
used to falsify this model.  Originally $\alpha_E=2$ was
suggested (Rossi et al. 2002, Zhang \& Meszaros 2002) for a
surface energy density that is a power law in the jet
off-axis angle, while a Gaussian surface energy density
would have $\alpha_E=1$ (Lloyd-Ronning et al. 2004).
Consequently our value of $\alpha_E=1.76\pm0.05$ falls in
the range of expected $\alpha_E$ values.  Thus our work
does not distinguish between the universal jet profile
model and the uniform jet model (which does not predict an
energy probability distribution).

Since the value of $\alpha_E$ is relevant to jet models,
consideration of the possible systematic errors is
warranted.  Our method for estimating the energy
probability distribution considers where each measured
$E_{\rm iso}$ falls between $E_{\rm iso,min}$ {\it for that
burst} and infinity.  If $E_{\rm iso,min}$ is
underestimated, then the low $E_{\rm iso}$ portion of the
probability distribution will be under-represented, causing
the distribution to be shifted to higher energy, in this
case resulting in a smaller value of $\alpha_E$.  Indeed,
when we use $P_{\rm ph,min}$=0.1, 0.3, 0.5, 1 and
2~ph~cm$^{-2}$~s$^{-1}$ we find $\alpha_E$=1.125, 1.6,
1.76, 1.9 and 2.  Next, uncertainties in the measured value
of $E_{\rm iso}$ that are symmetric to higher and lower
values will cause a net `diffusion' towards higher $E_{\rm
iso}$ because there are more bursts below than above any
given value of $E_{\rm iso}$.  This effect also decreases
$\alpha_E$.  Note that small lags are more difficult to
measure, resulting in greater uncertainties for large
$L_B$; since $E_{\rm iso}$ is correlated with $L_B$ (but is
not strictly proportional), on average the uncertainty
increases for larger $E_{\rm iso}$.  Therefore, the true
value of $\alpha_E$ is most likely greater than the value
we measured.

The likelihood alone does not indicate whether the power
law model is a good description of energy distribution. The
average of the cumulative probability (\S 2.3) should be
$\langle P(>E_{\rm iso})\rangle = 1/2 \pm (12N_B)^{-1/2}$
for $N_B$ bursts.  Figure~7 shows that the cumulative
$P(>E_{\rm iso})$ for $P_{\rm ph,min} =
0.5$~ph~cm$^{-2}$~s$^{-1}$ is very close to the straight
line that is expected if the distribution function is a
good description of the data. We calculate $\langle
P(>E_{\rm iso})\rangle = 0.4642$ with an expected
statistical variance of $0.0089$ ($N_B=1054$).  Thus
$\langle P(>E_{\rm iso})\rangle$ differs from the expected
value of 1/2 by $4\sigma$ using the statistical variance
only. However, since the likely systematic uncertainties
(e.g., in the value of $P_{\rm ph,min}$, in the
determination of the lags, in the corrections for the
redshifting of high energy light curves) are likely to be
considerable, this value of $\langle P(>E_{\rm
iso})\rangle$, and the proximity of the $P(>E_{\rm iso})$
distribution to a straight line (Figure~7), indicates that
a power law is a good representation of the distribution.

\subsubsection{Lognormal $E_{\rm iso}$ Distribution}

The second model distribution we consider is a lognormal
energy probability distribution; the parameters are the
energy centroid $E_{\rm iso,cen}$ and the logarithmic width
$\sigma_E$ of the distribution.  For this distribution the
likelihood (relevant to a frequentist analysis) and the
posterior distribution as a function of $\ln E_{\rm
iso,cen}$ and $\sigma_E$ (relevant to a Bayesian analysis)
are the same if the prior is constant in $\ln E_{\rm
iso,cen}$ and $\sigma_E$ (see Band 2001). Figure~8 shows
the likelihood surface; the peak occurs at $E_{\rm iso,cen}
= 2.8 \times 10^{50}$~ergs and $\sigma_E = 2.7$. However,
these parameters are highly correlated since a broader
distribution (larger $\sigma_E$) can compensate for a
smaller central energy $E_{\rm iso,cen}$.  The figure shows
that 95\% of the probability distribution is in the range
$E_{\rm iso,cen}=$(0.03--1)$\times10^{51}$~erg and
$\sigma_E=2.3$--3.2. As can be seen from figure~5, $E_{\rm
iso,cen} = 2.8 \times 10^{50}$~ergs is at the lower end of
the distribution of measured $E_{\rm iso,min}$, and thus
the data are insufficient to determine whether the
probability density does indeed decrease below $E_{\rm
iso,cen}$.

For $P_{\rm ph,min} = 0.5$~ph~cm$^{-2}$~s$^{-1}$, $\langle
P(>E_{\rm iso})\rangle = 0.4821 \pm 0.0089$ ($N_B=1054$),
which is consistent with $\langle P(>E_{\rm iso}) \rangle =
1/2$ at the $2\sigma$ level.  Note that for $P_{\rm ph,min}
= 0.3$~ph~cm$^{-2}$~s$^{-1}$ $\langle P(>E_{\rm
iso})\rangle = 0.4598 \pm 0.0085$ ($N_B=1162$), which
differs from the expected value by nearly $5\sigma$ using
the statistical variance only.  As we argued above, there
are undoubtedly significant systematic uncertainties in
addition to the statistical variance.  Figure~9 shows the
distribution of the cumulative probability for $P_{\rm
ph,min} = 0.5$~ph~cm$^{-2}$~s$^{-1}$; as can seen, the
observed cumulative probability distribution is very close
to the expected distribution.  Again, considering the
systematic uncertainties in our calculation, the lognormal
distribution is consistent with the data.

\subsubsection{Comparison to Previous Probability Distribution Calculations}

In this study we apply the same methodology presented in
Band (2001) to a new burst data set.  In both studies the
$E_{\rm iso}$ distributions are modelled with lognormal and
power law functional forms, although Band (2001) applied a
high energy cutoff to the power law function because in
some cases $\alpha_E\le 1$.  Band (2001) used three data
sets: a)~`B9'---9 bursts with spectroscopic redshifts
(i.e., the redshifts were measured from emission or
absorption lines) and fitted BATSE spectra; b)~`C17'---17
bursts with spectroscopic redshifts and spectral
information from a variety of sources; and c)~`F220'---220
bursts with redshifts derived from the
variability-luminosity relation (Fenimore \& Ramirez-Ruiz
2000). Table~1 presents the resulting parameters; 90\%
uncertainties are indicated. The variables of the lognormal
distribution are highly correlated, resulting is large
uncertainty ranges for both parameters. We find that as the
database size increases, it samples smaller $E_{\rm iso}$,
as can be seen from the values of $E_2$. Perhaps as a
consequence, $E_{\rm iso,cen}$ decreases and $\alpha_E$
increases as the database size increases.

The distributions for these different samples are
discrepant, which may result from systematic difficulties
with the burst samples or incorrect assumed probability
distribution functional forms. The variability-luminosity
relation used for the F220 sample and the lag-luminosity
relation used in this paper were both calibrated with only
a few bursts, and the validity of these relations must be
confirmed by a larger burst sample.  The B9 and C17 samples
use spectroscopic redshifts, and thus are affected by
selection effects in detecting such redshifts.  Band (2001)
used the $P_{\rm ph,min}$ for the BATSE detection, but the
true $P_{\rm ph,min}$ for measuring the redshift was
undoubtedly significantly greater:  the bursts in this
sample were usually detected and rapidly localized by a
less sensitive detector (e.g., {\it Beppo-SAX}); and the
burst's afterglow had to be sufficiently bright for an
improved localization that warranted follow-up
spectroscopic observations. Thus the B9 and C17 samples
were seriously flawed.  Note that the discrepancy between
the F220 sample and ours is not very great, particularly
for the power law distribution.

The systematic trends in the fitted parameters of the
probability distribution function with the burst sample's
$E_{\rm iso}$ range suggest that a simple power law or a
lognormal form are not the correct functional form. For
example, Schaefer, Deng, \& Band (2001) and Lloyd-Ronning,
Fryer \& Ramirez-Ruiz (2002) find that the luminosity
function for $L_B$ is a broken power law.

Comparison to other studies is more difficult, particularly
since we study the probability distribution of $E_{\rm
iso}$ while others (Fenimore \& Ramirez-Ruiz 2000; Schaefer
et al. 2001; Schmidt 2001; Lloyd-Ronning et al. 2002;
Norris 2002) studied the luminosity function of $L_B$.  In
addition, these studies use different definitions of the
luminosity, e.g., for BATSE's 50-300~keV trigger band
(Fenimore \& Ramirez-Ruiz 2000; Lloyd-Ronning 2002) or
bolometric ($E=0.025$--20~MeV for Norris 2002), with
transformations using a variety of spectra (e.g., an
$E^{-2}$ power law for Schaefer et al. 2001 and Schmidt
2001, a broken power law for Norris 2002 and a `Band'
spectrum with fixed parameters for all bursts for Fenimore
\& Ramirez-Ruiz 2000).  Consequently, a more detailed
comparison between distributions is beyond the scope of our
study.

\section{Summary}

In this paper we have two objectives.  First, we use the
lag-luminosity relation to calculate self-consistently the
redshifts for 1218 BATSE bursts. For the bursts without the
spectral parameters required by the calculation we use
average low and high energy spectral indices, and a peak
energy $E_p$ derived from the hardness ratio.  We find that
the redshift is quite sensitive to the spectral parameters,
particularly if the spectrum is sharply peaked. We use both
single power law and broken power law lag-luminosity
relations, and find that the broken power law relation does
indeed predict a population of low luminosity, nearby
bursts.  For both forms of the relation the median redshift
is 1.58.

We use the redshifts to calculate the apparent peak
bolometric flux and the isotropic energy, both assuming
that the bursts radiate isotropically.

Second, we fit two functional forms to the distribution of
the isotropic energy.  We find that our burst data can be
fit by a power law energy distribution with
$\alpha_E=1.76\pm0.05$ (95\% confidence); considering the
likely systematic uncertainties in addition to the
statistical variance, the power law distribution is
probably a good description of the data.  This value of
$\alpha_E$ is in the acceptable range for the universal jet
profile model, and therefore, our work does not distinguish
between the current jet structure models.  A lognormal
energy distribution also describes the data; the data
permit a smaller average energy if the distribution is
wider.  More faint bursts will bound the lower end of this
distribution.

\acknowledgements
We thank the referee for helpful comments
on improving the text of our paper.  We also thank
V.~Avila-Reese for pointing out an error in the text.

\clearpage

\begin{figure}
\plotone{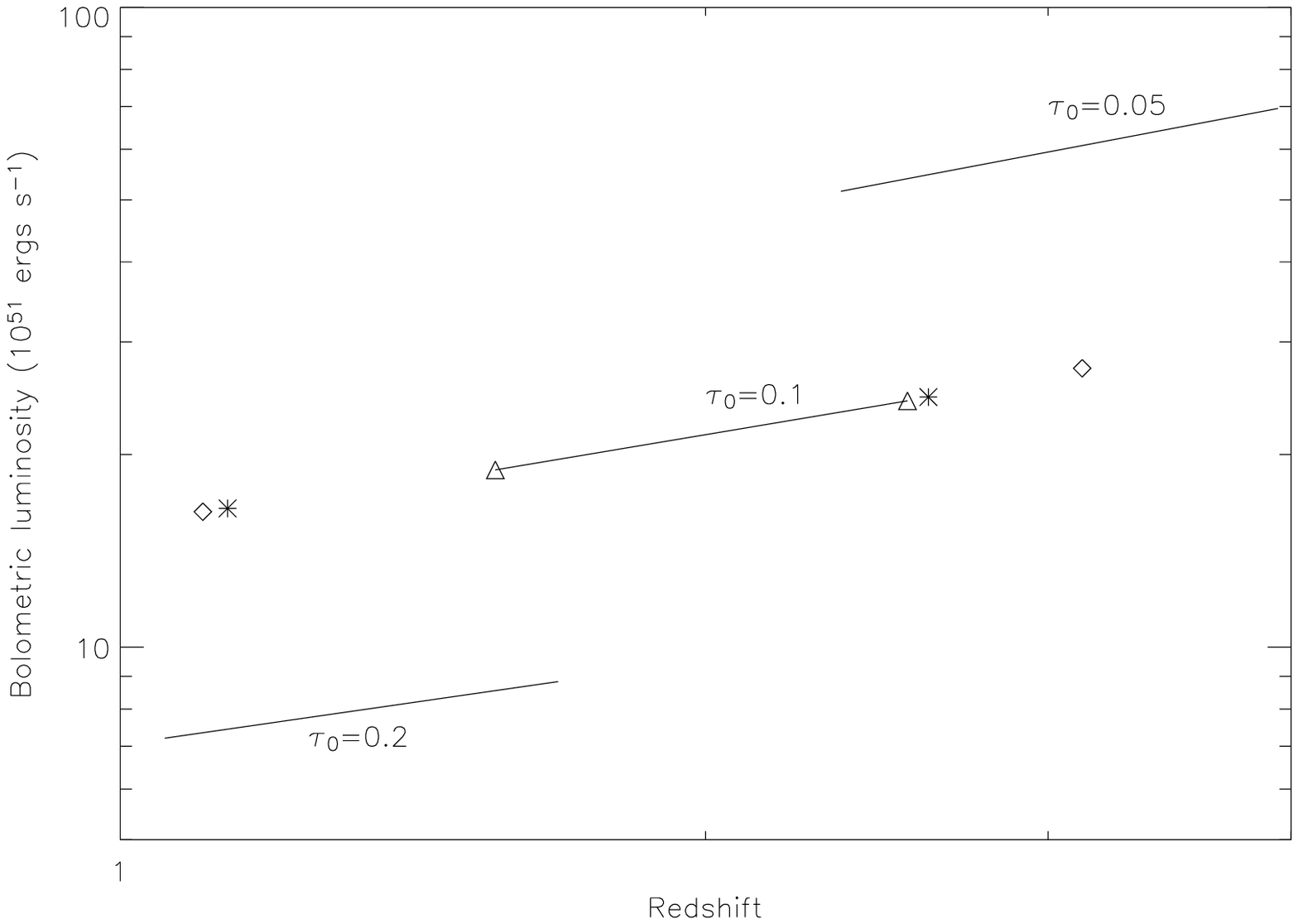} \caption{Variation of bolometric peak
luminosity $L_B$ as a function of redshift as $E_p$ varies.
The solid curves, labelled by the observed lag $\tau_0$,
are for $\alpha=-1$ and $\beta=-2$ while $E_p$ varies
between 5~keV and 2~MeV. Also shown on the $\tau_0=0.1$
curve and its extrapolation are the maximum and minimum
redshifts for: triangle---$\alpha=-1$, $\beta=-2$;
diamond---$\alpha=-1$, $\beta=-3$; and
asterisk---$\alpha=-1/2$, $\beta=-2$.  In all cases $P_{\rm
ph}= 1$~ph~cm$^{-2}$~s$^{-1}$. \label{z_l_tau}}
\end{figure}

\begin{figure}
\plotone{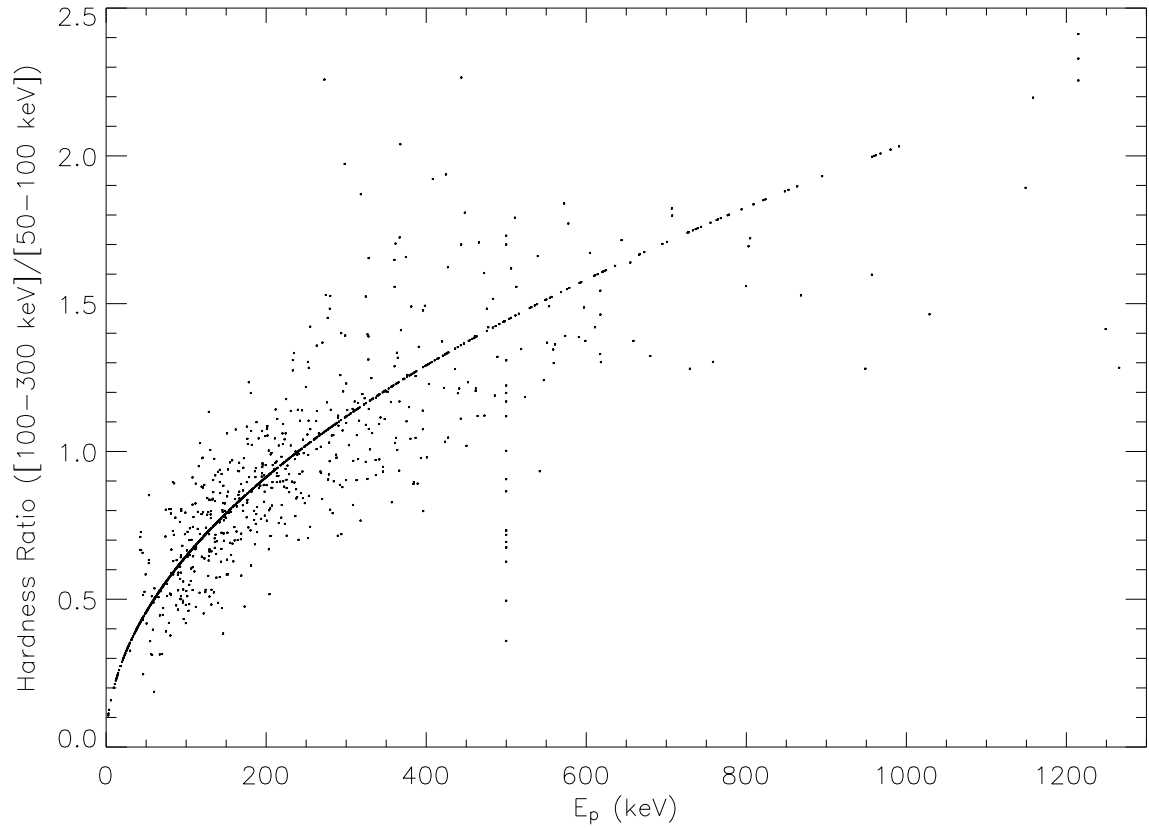} \caption{Hardness ratio HR$_{32}$
(100--300~keV vs. 50--100~keV) as a function of $E_p$.  Two
populations are evident.  First, spectral fits were not
available for the bursts that fall on an empirical
$E_p\propto \hbox{HR}_{32}^2$ relation. Second, bursts with
spectral fits are dispersed around this $E_p \propto
\hbox{HR}_{32}^2$ relation. \label{ep_hr32}}
\end{figure}

\begin{figure}
\plotone{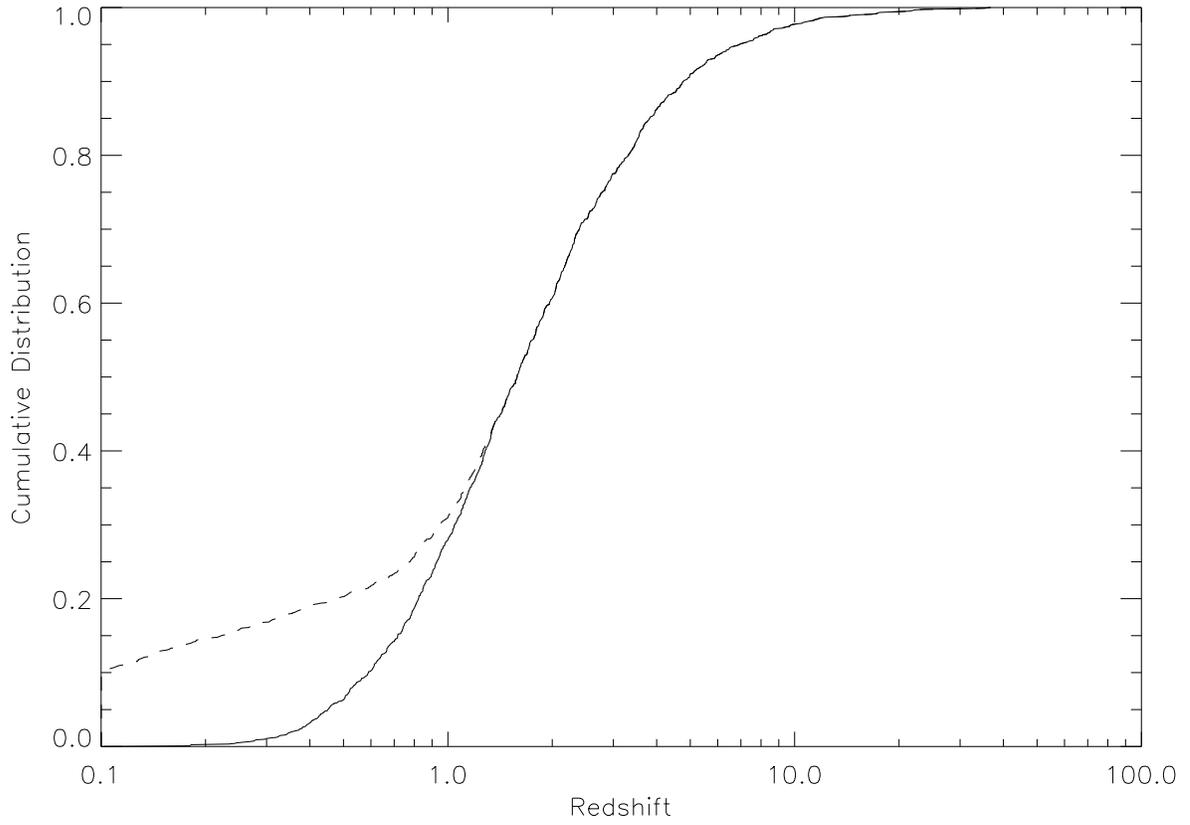} \caption{Cumulative distribution of
bursts by redshift.  Solid line---one component
lag-luminosity relation.  Dashed line---two component
relation.  Both distributions have the same median redshift
of 1.6. \label{z_cum}}
\end{figure}

\begin{figure}
\plotone{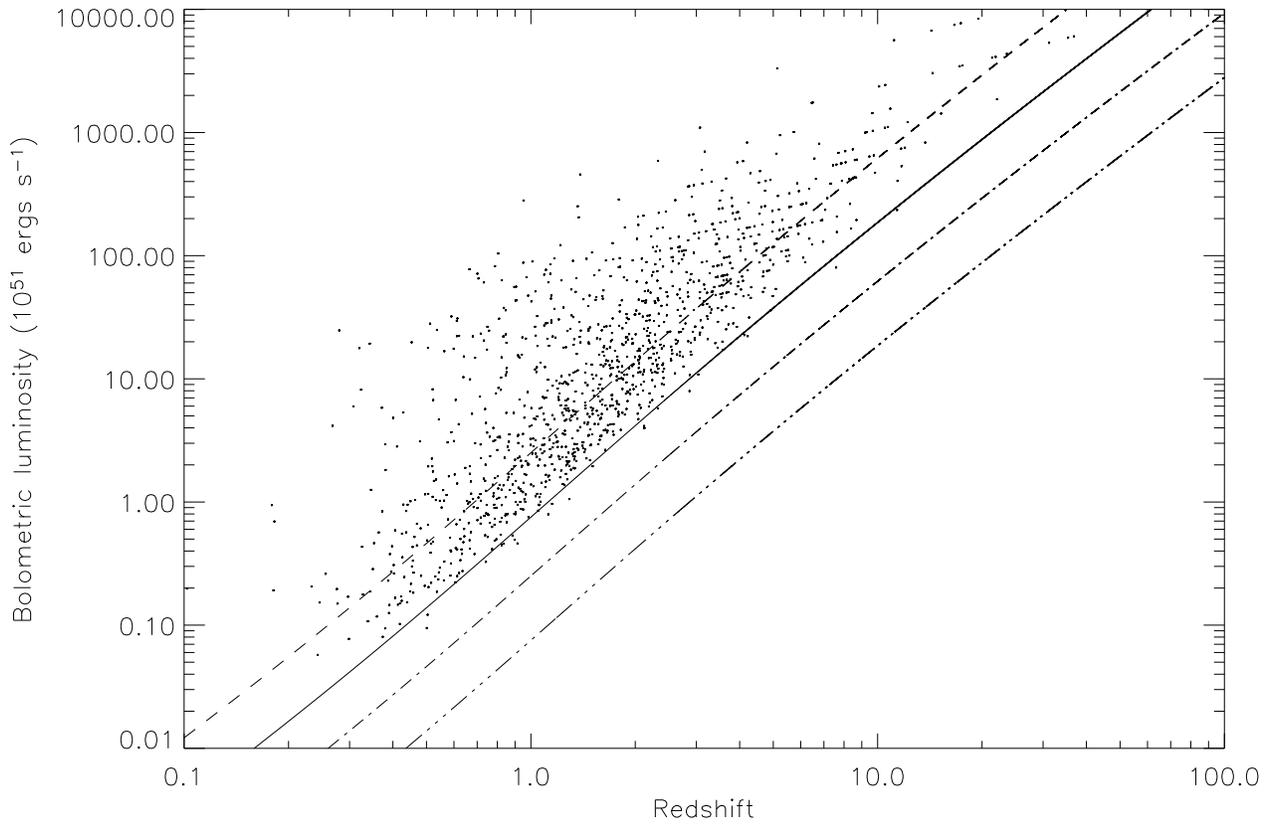} \caption{Bolometric peak
luminosity $L_B$ vs. redshift $z$ for the single component
lag-luminosity relation.  The lines represent the threshold
peak luminosity for a threshold peak flux of $P_{\rm
ph,min} = 0.3$~ph~cm$^{-2}$~s$^{-1}$ in the 50--300~keV
band, $\alpha=-1$, $\beta=-2$, and $\langle E\rangle =30$
(2 dots-dashed line), 100 (dot-dashed), 300 (solid) and
1000~keV (dashed), where $\langle E \rangle$ is defined by
eq.~4. \label{z_l_1comp_spec2}}
\end{figure}

\begin{figure}
\plotone{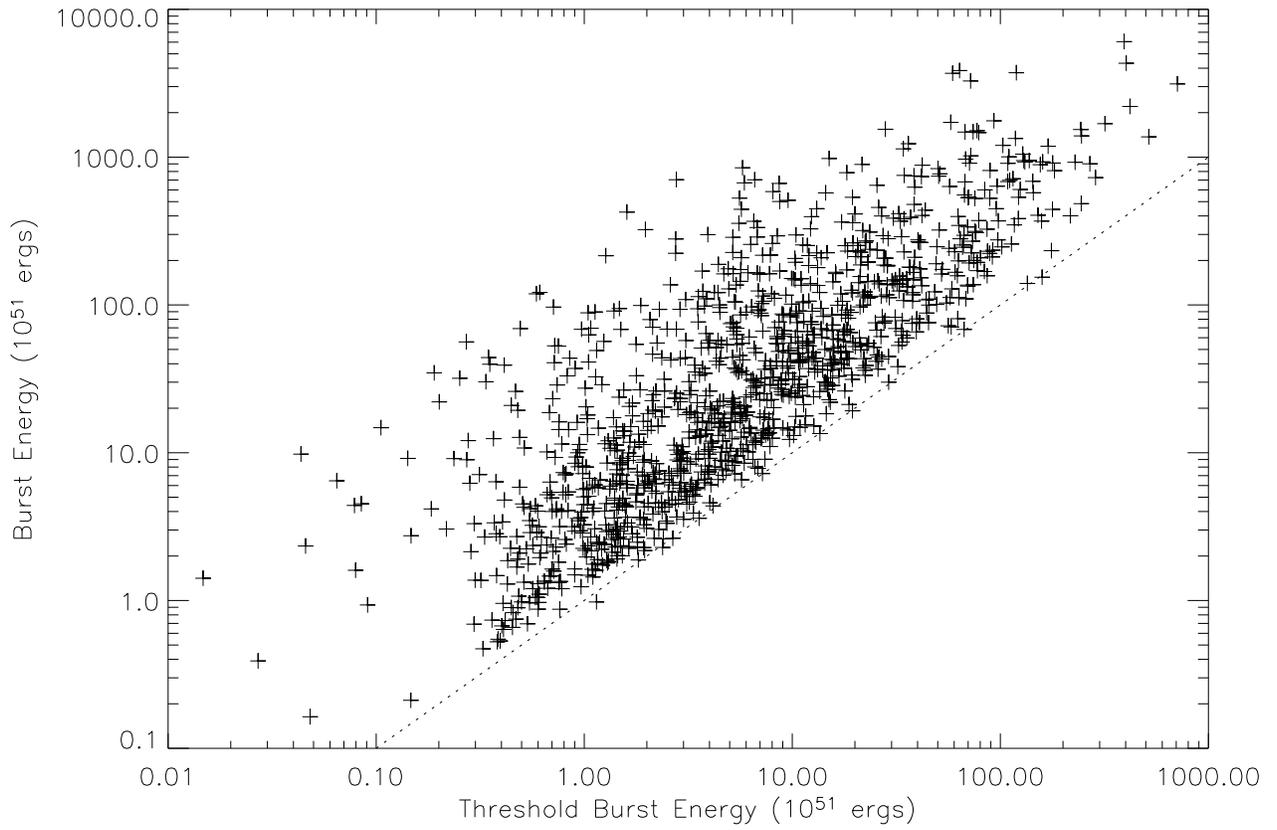} \caption{Scatter plot of $E_{\rm iso}$
vs. the $E_{\rm iso,min}$.  On the dotted line $E_{\rm
iso}=E_{\rm iso,min}$.  The paucity of bursts just above
the dotted line suggests that $P_{\rm ph,min}$ should be
raised from 0.3 to 0.5~ph~cm$^{-2}$~s$^{-1}$.
\label{e_scat}}
\end{figure}

\begin{figure}
\plotone{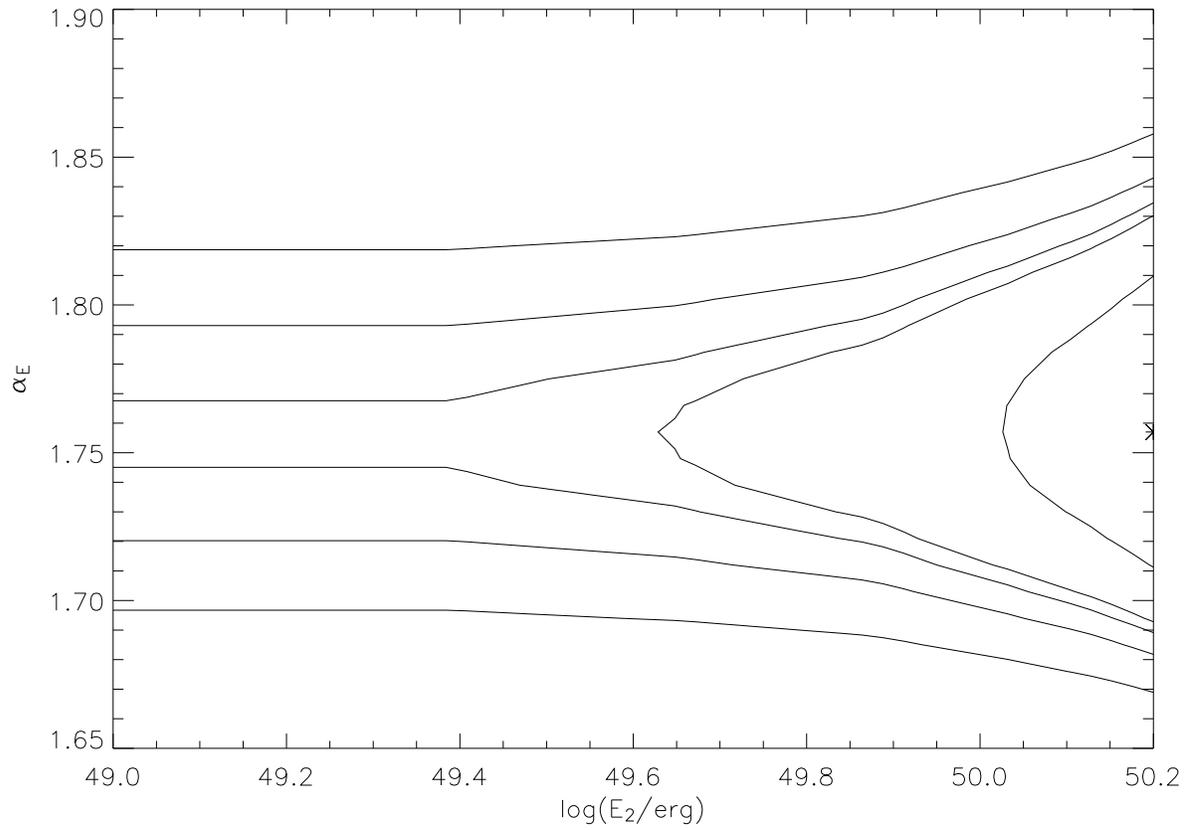} \caption{Contour plot of the
likelihood function for a power law energy probability
distribution. A threshold peak flux of $P_{\rm ph,min} =
0.5$~ph~cm$^{-2}$~s$^{-1}$ was imposed.  The power law has
an index of $\alpha_E$ and a low energy cutoff of $E_2$.
The contour levels contain 0.68, 0.9, 0.95, 0.99 and 0.999
of the integrated probability. \label{pl_normalized_p5}}
\end{figure}

\begin{figure}
\plotone{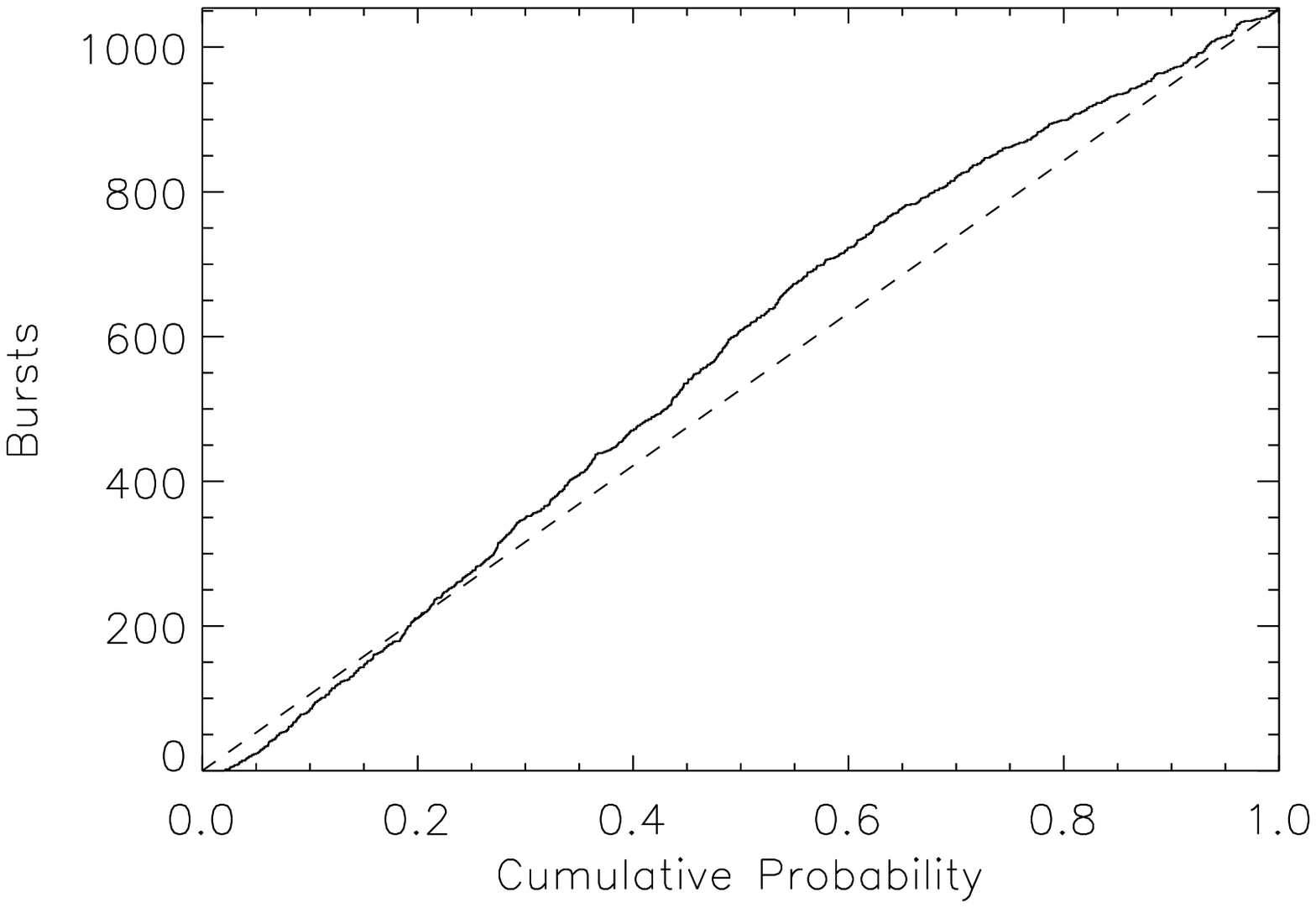} \caption{Cumulative
distribution of the probability of the observed isotropic
energies for a power law energy probability distribution.
A threshold peak flux of $P_{\rm ph,min} =
0.5$~ph~cm$^{-2}$~s$^{-1}$ was imposed.
\label{pl_cum_prob_p5}}
\end{figure}

\begin{figure}
\plotone{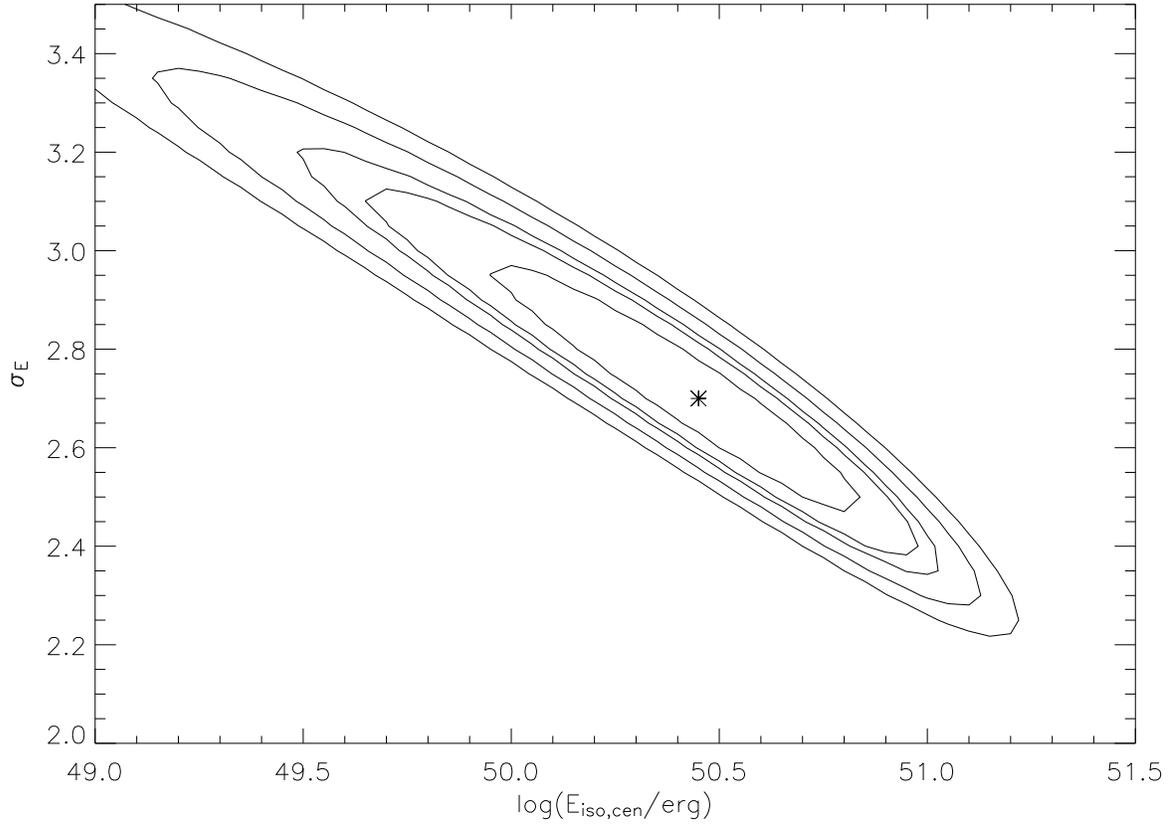} \caption{Contour plot of the
likelihood function for a lognormal energy probability
distribution.  A threshold peak flux of $P_{\rm ph,min} =
0.5$~ph~cm$^{-2}$~s$^{-1}$ was imposed.  The distribution
is characterized by a centroid energy and a logarithmic
width. The contour levels contain 0.68, 0.9, 0.95, 0.99 and
0.999 of the integrated probability.
\label{ln_normalized_p5}}
\end{figure}

\begin{figure}
\plotone{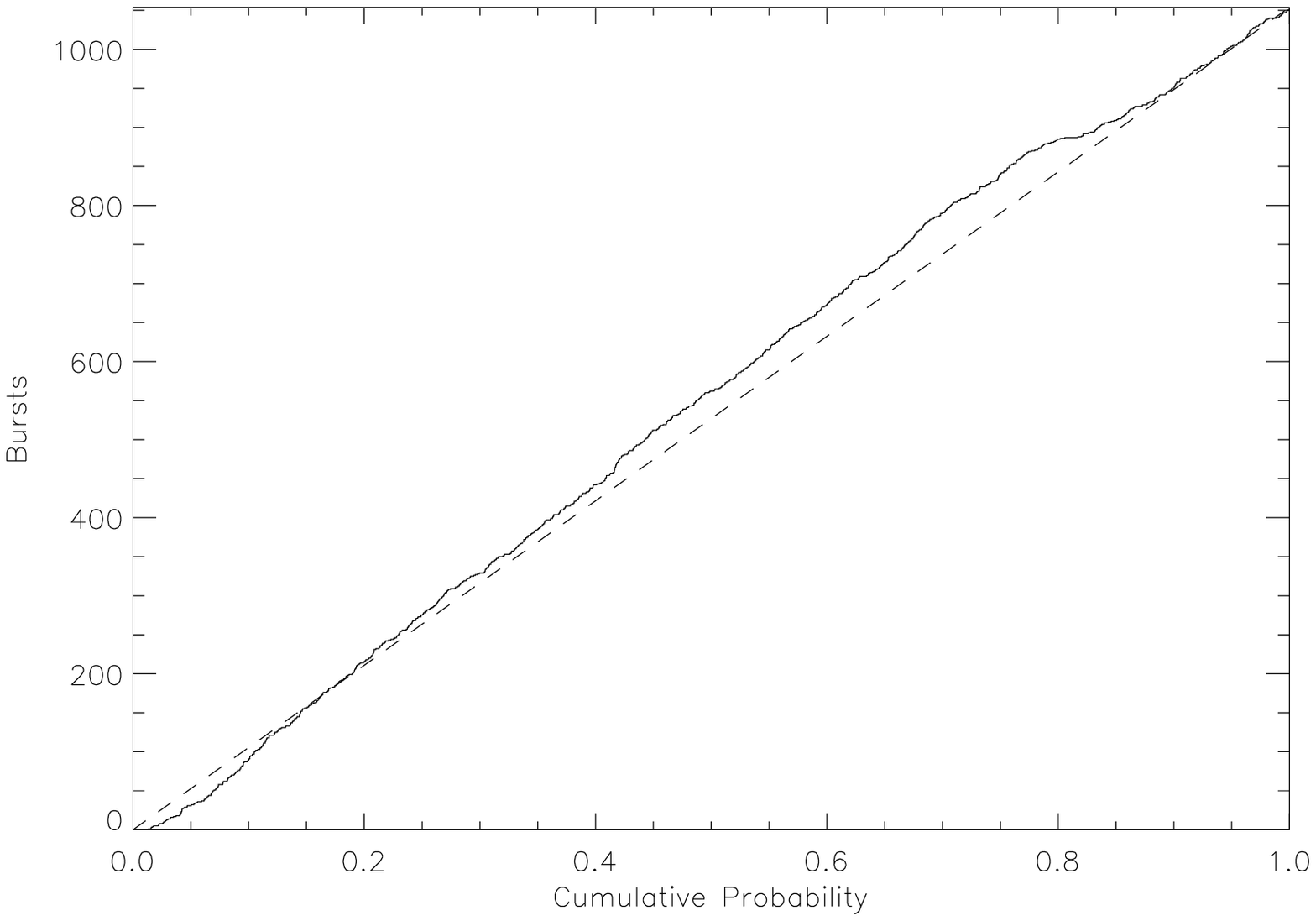} \caption{Cumulative distribution of
the probability of the observed isotropic energies for a
lognormal energy probability distribution.  A threshold
peak flux of $P_{\rm ph,min} = 0.5$~ph~cm$^{-2}$~s$^{-1}$
was imposed. \label{p_dist_p5}}
\end{figure}

\clearpage

\begin{deluxetable}{l c c c c}
\tablecolumns{5}
%\tablewidth{0in}
\tablecaption{\label{Table1}Probability
Distribution Parameters for Different Burst Samples}
\tablehead{
\colhead{Quantity \qquad\qquad} &
\colhead{\qquad B9\tablenotemark{a}\qquad\qquad} &
\colhead{\qquad C17\tablenotemark{b}\qquad\qquad} &
\colhead{\qquad F220\tablenotemark{c}\qquad\qquad} &
\colhead{\qquad LL1054\tablenotemark{d}\qquad}}
\startdata
$E_{\rm iso,cen}$\tablenotemark{e} & $130^{+190}_{-124}$ &
$52^{+48}_{-50}$ & $12^{+11}_{-10}$ &
$0.28^{+0.61}_{-0.23}$
\\
$\sigma_E$\tablenotemark{f} & $1.9^{+2.6}_{-0.4}$ &
$2.1^{+2.1}_{-0.4}$ & $1.9^{+0.4}_{-0.2}$ &
$2.7^{+0.4}_{-0.3}$ \\
$\alpha_E$\tablenotemark{g} & $0.74^{+0.46}_{-0.34}$ &
$0.96^{+0.29}_{-0.21}$ & $1.81^{+0.13}_{-0.11}$ &
$1.76^{+0.07}_{-0.07}$ \\
$E_2$\tablenotemark{h} & 1.6 & 0.55 & 0.12 & 0.16 \\
$E_c$\tablenotemark{i} & 1440 & 1460 & 5000 & --- \\
\enddata
\tablenotetext{a}{Sample of 9 BATSE bursts with
spectroscopic redshifts and fitted spectra; analyzed
in Band (2001). The threshold $E_{\rm iso,min}$ was
significantly underestimated.}
\tablenotetext{b}{Sample of 17 bursts with spectroscopic
redshifts; analyzed in Band (2001).  The threshold
$E_{\rm iso,min}$ was significantly underestimated.}
\tablenotetext{c}{Sample of 220 bursts with redshifts
derived from variability redshifts; analyzed in Band
(2001).}
\tablenotetext{d}{The sample used in this work, with
$P_{\rm ph,min} = 0.5$~ph~cm$^{-2}$~s$^{-1}$.}
\tablenotetext{e}{The central energy of the lognormal
distribution, in units of $10^{51}$~erg.  }
\tablenotetext{f}{The logarithmic width (in units of the
energy's natural logarithm) for the lognormal distribution.}
\tablenotetext{g}{The index of the power law
distribution, $p(E_{\rm iso}) \propto E_{\rm
iso}^{-\alpha_E}$. }
\tablenotetext{h}{The low energy cutoff of the power law
distribution, in units of $10^{51}$~erg.  This energy is
the lowest $E_{\rm iso,min}$ for the sample.}
\tablenotetext{i}{The high energy cutoff of the power law
distribution, in units of $10^{51}$~erg.}

\end{deluxetable}


\clearpage

\begin{thebibliography}{}

\bibitem[a1]{a1}Band, D. 2001, ApJ, 563, 582

\bibitem[a5]{a5}Band,~D., et al. 1993, ApJ, 413, 281

\bibitem[a12]{a12}Bloom, J. S., Frail, D., \& Kulkarni, S.
2003, ApJ, 594, 674

\bibitem[a114]{a114}Fenimore, E. E., in't Zand, J. J. J. M.,
Norris,~J.~P., Bonnell,~J.~T., \& Nemiroff,~R.~J. 1995,
ApJ, 448, L101

\bibitem[a13]{a13}Fenimore, E. E., \& Ramirez-Ruiz, E.
2000, ApJ, submitted [astro-ph/0004176]

\bibitem[a11]{a11}Frail, D., et al. 2001, ApJ, 562, 55

\bibitem[1]{1}Lloyd-Ronning, N., Dai, X., \& Zhang, B. 2003,
ApJ, submitted [astro-ph/0310431]

\bibitem[a111]{a111}Lloyd-Ronning, N., Fryer, C. L., \& Ramirez-Ruiz,~E.
2002, ApJ, 574, 554

\bibitem[6]{6}Mallozzi, R., et al. 1998, in Gamma-Ray
Bursts, 4th Huntsville Symposium, AIP Conference
Proceedings 428, eds. C. Meegan, R. Preece and T. Koshut
(AIP: Woodbury, NY), 273

\bibitem[a2]{a2}Norris, J. P. 2002, ApJ, 579, 386

\bibitem[a4]{a4}Norris, J. P., Marani, G. F., \& Bonnell,
J. T. 2000, ApJ, 534, 248

\bibitem[a3]{a3}Preece,~R.~D., Briggs,~M.~S.,
Mallozzi,~R.~S., Pendleton,~G.~N., Paciesas,~W.~S., \&
Band,~D.~L. 2000, ApJS, 126, 19

\bibitem[a14]{a14}Reichart, D. E., Lamb, D. Q., Fenimore, E. E., Ramirez-Ruiz,
E., Cline, T. L., \& Hurley, K. 2001, ApJ, 552, 57

\bibitem[a7]{a7}Rossi, E., Lazzati, D., \& Rees,~M.~J.
2002, MNRAS, 332, 945

\bibitem[3]{3}Salmonson, J. 2001, ApJ, 546, 29

\bibitem[a42]{a42}Schaefer, B. E., Deng, M., \& Band, D. L.
2001, 563, L123

\bibitem[a41]{a41}Schmidt, M. 2001, ApJ, 552, 36

\bibitem[a8]{a8}Zhang, B., \& Meszaros, P. 2002, ApJ, 571,
876

\bibitem[a9]{a9}Zhang, W., Woosley, S. E., \&
MacFayden,~A.~I. 2003, ApJ, 586, 356

\end{thebibliography}
\end{document}